\documentstyle[preprint,aps]{revtex}

\newcommand{\be}{\begin{equation}}
\newcommand{\ee}{\end{equation}}
\newcommand{\bea}{\begin{eqnarray}}
\newcommand{\eea}{\end{eqnarray}}
\newcommand{\ba}{\begin{array}}
\newcommand{\ea}{\end{array}}
\newcommand{\nn}{\nonumber \\}

\begin{document} 
\tightenlines

\preprint{}

\title{\vspace{2cm} \Large \bf NONCOMMUTATIVE THEORIES, \\ SEIBERG-WITTEN MAP
  AND GRAVITY}

\author{Victor O. Rivelles}
\address{Instituto de F\'{\i}sica \\ Universidade de S\~{a}o Paulo\\
 Caixa Postal 66318, 05315-970, S\~{a}o Paulo, SP, Brazil\\
E-mail: rivelles@fma.if.usp.br}

\maketitle
\thispagestyle{empty}

\begin{abstract}
We review the connection between noncommutative field
theories and gravity. When the noncommutativity is induced by the
Moyal product we can use the Seiberg-Witten map in order to deal with
ordinary fields. We then show that the effect of the noncommutativity
is the same as a field dependent gravitational background. The
gravitational background is that of a gravitational plane wave and the
coupling is charge dependent. Uncharged fields couple more
strongly than the charged ones. Deviations from the usual dispersion
relations are discussed and we show that they are also charge
dependent. 
\end{abstract}

\vspace{1cm}

Noncommutative (NC) theories have a long history
\cite{Szabo:2001kg}. More recently they 
were found in a limit of string theory with D-branes in a constant 
NS-NS background B field \cite{Seiberg:1999vs}. In this limit gravity
decouples but still leaves a trace in the emerging NC gauge theory. If
point splitting regularization is used in the world-sheet then NC
appears through the Moyal product which is defined as 
\be
\label{Moyal}
A(x) \star B(x) = e^{\frac{i}{2} \theta^{\mu\nu} \partial^x_\mu
  \partial^y_\nu} A(x) B(y)|_{y\rightarrow x},
\ee
where $\theta^{\mu\nu}$ is the NC parameter. 
If Pauli-Villars regularization is used instead then ordinary gauge
symmetry is preserved and NC appears in higher dimension
operators. Since the S matrix must be regularization independent these
two descriptions are related by a space-time field redefinition
known as the Seiberg-Witten (SW) map \cite{Seiberg:1999vs}. 

From the space-time point of view these two descriptions have quite
different properties. When the Moyal product is used the ultraviolet
structure of the theory is not modified \cite{Filk:dm} but new
infrared divergences appear and get mixed with the ultraviolet ones
\cite{Minwalla:1999px}. This mixing of divergences turns the theory
non-renormalizable except in some cases where supersymmetry is
present \cite{Girotti:2000gc}. An important property of NC theories
induced by the Moyal product, which distinguishes them from 
the conventional ones, is that translations in the NC directions are
equivalent to gauge transformations \cite{Gross:2000ph}. This can be
seen even for the case of a scalar field which has the NC gauge
transformation 
$ \delta \hat{\phi} = -i [ \hat{\phi}, \hat{\lambda} ]_\star$,
where $[A,B]_\star  = A \star B - B \star A$ is the Moyal
commutator. Under a global translation the scalar field transforms as
$\delta_T \phi = \xi^\mu \partial_\mu \hat{\phi}$. Derivatives of the
field can be rewritten using the Moyal commutator as $\partial_\mu
\hat{\phi} = - i \theta^{-1}_{\mu\nu} [ x^\nu, \hat{\phi}]_\star$ so
that $\delta \hat{\phi} = \delta_T \hat{\phi}$ with gauge parameter
$\hat{\lambda} = - \theta^{-1}_{\mu\nu} \xi ^\mu x^\nu$. The only other
field theory which has a similar property is general relativity where
local translations are gauge transformations associated to general
coordinate transformations. This remarkable property shows that, as in
general relativity, there are no local gauge invariant observables in
NC theories.

In the description obtained through the SW map the theory is 
presented as a series expansion in $\theta$. In this
way a local field theory is obtained at the expense of introducing
a large number of non-renormalizable interactions
\cite{Wulkenhaar:2002ps}.  At the classical level, on the other 
side, it is possible to understand very clearly the breakdown of
Lorentz invariance induced by the noncommutativity. The dispersion
relation for plane waves in a magnetic background gets modified so
that photons do not move with the velocity of light
\cite{Guralnik:2001ax}. However, the connection between
translations and gauge transformations seems to be lost. A global
translation on a commutative real scalar field $\delta_T \phi = \xi^\mu
\partial_\mu {\phi}$ can no no longer be rewritten as a gauge
transformation since $\delta \phi = 0$. We will discuss how other
aspects concerning gravity emerges in the description which makes use
of the SW map. In this case NC field theories can be interpreted as
ordinary theories immersed in a gravitational background generated by
the gauge field. We will show that the
$\theta$ dependent terms in the commutative action can be interpreted
as a gravitational background which depends on the gauge field. 
We then determine the metric which couples to real and
complex scalar fields. We find that the uncharged field coupling is
twice that of the charged one. So we can interpret the gauge coupling
in NC theory as a particular gravitational coupling which depends on
the charge of the field. We also determine the geodesics
followed by  a massless particle in this background. We find that its
velocity differs from the velocity of light by an amount proportional
to $\theta$ with the deviation for the uncharged case being twice that
of the charged one. For the uncharged case the deviation is the same
as that found for the the gauge theory in flat space-time
\cite{Guralnik:2001ax,Cai:2001az}. As a final check we derive these
same velocities in a field theoretic context.

The action for the NC Abelian gauge theory in flat space-time is
\be
\label{NC_gauge_action}
S_A = -\frac{1}{4} \int d^4x \,\,\, \hat{F}^{\mu\nu} \star
\hat{F}_{\mu\nu},
\ee
where $\hat{F}_{\mu\nu} = \partial_\mu \hat{A}_\nu - \partial_\nu
\hat{A}_\mu - i [ \hat{A}_\mu, \hat{A}_\nu ]_\star$. For
a real scalar field in the adjoint representation of $U(1)$ the flat
space-time action is
\be
\label{NC_scalar_action}
S_\varphi = \frac{1}{2} \int d^4x \,\,\, \hat{D}^\mu \hat{\varphi} \star
\hat{D}_\mu \hat{\varphi},
\ee
where $\hat{D}_\mu \hat{\varphi} = \partial_\mu \hat{\varphi} - i
[\hat{A}_\mu, \hat{\varphi} ]_\star$.
On the other side, for a complex scalar field in the fundamental
representation of $U(1)$ the action is
\be
\label{NC_complex_scalar_action}
S_\phi = \int d^4 x \,\,\, \hat{D}^\mu \hat{\phi} \star ( \hat{D}_\mu
\hat{\phi} )^\dagger,
\ee
with $\hat{D}_\mu \hat{\phi} = \partial_\mu \hat{\phi} - i \hat{A}_\mu
\star \hat{\phi}$. The gauge transformations which leave the above
actions invariant are given by
\be
\label{gauge_transformations}
\delta \hat{A}_\mu = \hat{D}_\mu \hat{\lambda}, \qquad
\delta \hat{\varphi} = - i [ \hat{\varphi}, \hat{\lambda} ]_\star,
\qquad \delta \hat{\phi} = i
\hat{\lambda} \star \hat{\phi}, \qquad \delta \hat{\phi}^\dagger = -i
\hat{\phi}^\dagger \star \hat{\lambda}.
\ee

To first order in $\theta$ the SW map is
\cite{Seiberg:1999vs,Rivelles:2002ez} 
\bea
\label{SW_map}
\hat{A}_\mu &=& A_\mu - \frac{1}{2} \theta^{\alpha\beta}
A_\alpha ( \partial_\beta A_\mu + F_{\beta\mu} ), \nn
\hat{\varphi} &=& \varphi - \theta^{\alpha\beta} A_\alpha
\partial_\beta \varphi, \nn
\hat{\phi} &=& \phi - \frac{1}{2} \theta^{\alpha\beta} A_\alpha
\partial_\beta \phi,
\eea
We can now expand the NC actions
(\ref{NC_gauge_action}),(\ref{NC_scalar_action}) and
(\ref{NC_complex_scalar_action}) using (\ref{Moyal}) and apply the map
(\ref{SW_map}) to get the corresponding commutative actions.

For the real scalar field we find, always to first order in
$\theta$,
\be
\label{action_scalar_field}
S_\varphi = \frac{1}{2} \int d^4x \, \left[ \partial^\mu \varphi \partial_\mu
\varphi + 2 \theta^{\mu\alpha} {F_\alpha}^\nu \left( - \partial_\mu \varphi
\partial_\nu \varphi + \frac{1}{4} \eta_{\mu\nu} \partial^\rho \varphi
\partial_\rho \varphi \right) \right].
\ee
It is worth to remark that the tensor inside the parenthesis is
traceless. If we now consider this same field coupled to a
gravitational background
\be
\label{action_scalar_gravity}
S_{g,\varphi} = \frac{1}{2} \int d^4x \, \sqrt{-g} g^{\mu\nu}
\partial_\mu \varphi \partial_\nu \varphi,
\ee
and expand the metric $g_{\mu\nu}$ around the flat metric
$\eta_{\mu\nu}$,
\be
g_{\mu\nu} = \eta_{\mu\nu} + h_{\mu\nu} + \eta_{\mu\nu} h,
\ee
where $h_{\mu\nu}$ is traceless, we get
\be
\label{linearized_scalar_action}
S_{g,\varphi} = \frac{1}{2} \int d^4x \, \left( \partial^\mu \varphi
\partial_\mu \varphi - h^{\mu\nu}  \partial_\mu \varphi \partial_\nu
\varphi + h \partial^\rho \varphi \partial_\rho \varphi \right),
\ee
where indices are raised and lowered with the flat metric.
Since both actions, (\ref{action_scalar_field}) and
(\ref{linearized_scalar_action}), have the same structure we can
identify a linearized background gravitational field
\bea
\label{scalar_linearized_metric}
h^{\mu\nu} &= & \theta^{\mu\alpha} {F_\alpha}^\nu + \theta^{\nu \alpha}
{F_\alpha}^\mu + \frac{1}{2} \eta^{\mu\nu} \theta^{\alpha\beta}
F_{\alpha\beta},\nn
h &=& 0.
\eea
Then, the effect of noncommutativity on the commutative scalar field
is similar to a field dependent gravitational field
\cite{Rivelles:2002ez}.

The same procedure can be repeated for the complex scalar field
\cite{Rivelles:2002ez}. We find the linearized metric  
\bea
\label{linearized_metric_complex}
h^{\mu\nu} &=& \frac{1}{2} \left( \theta^{\mu\alpha} {F_\alpha}^\nu +
\theta^{\nu\alpha} {F_\alpha}^\mu \right) + \frac{1}{4} \eta^{\mu\nu}
\theta^{\alpha\beta} F_{\alpha\beta}, \nn
h &=& 0.
\eea
Then charged fields feel a gravitational background which is half of
that felt by the uncharged ones. Therefore, the gravity coupling is now
dependent on the charge of the field, being stronger for uncharged
fields. 

Notice that the gauge field has now a dual role, it couples
minimally to the charged field and also as a gravitational
background. As it is well known the SW map
gives rise to the following action
\be
\label{SW_gauge_action}
S_A = -\frac{1}{4} \int d^4x \,\,\, \left[ F^{\mu\nu} F_{\mu\nu} + 2
\theta^{\mu\rho} {F_\rho}^\nu \left( {F_\mu}^\sigma F_{\sigma\nu} +
\frac{1}{4} \eta_{\mu\nu} F^{\alpha\beta} F_{\alpha\beta} \right)
\right].
\ee
Again, the tensor inside the parenthesis is traceless. At this point
we could be tempted to consider this action as some gravitational
action build up from the metric (\ref{scalar_linearized_metric}) or
(\ref{linearized_metric_complex}). Since the field strength always
appears multiplied by $\theta$ inside the metric, all invariants
constructed with it will be of order $\theta$. Hence, they can not
give rise to (\ref{SW_gauge_action}), unless they appear in
combinations involving the inverse of $\theta$. If we insist in having
an action which is polynomial in $\theta$ 
the best we can do is to regard the gauge field as having
a double role again and couple it to gravitation as in the previous
case. We then find that $h^{\mu\nu}$ is given by
  (\ref{linearized_metric_complex}). Since the NC gauge field
resembles a non-Abelian gauge field we expect that its commutative
counterpart couple to the same gravitational field as the charged
one. It should also be remarked that in this case the
gravitational field can not be interpreted just as a fixed
background since it depends on the dynamical gauge field.

Having determined the field dependent background metric we can now
study its properties. A detailed analysis shows that it describes a
plane gravitational wave \cite{Rivelles:2002ez}. 

We can now turn our attention to the behavior of a massless particle
in this background. Its geodesics is described by
\be
\label{ds}
ds^2 = \left( 1 + \frac{1}{4} \theta^{\alpha\beta} F_{\alpha\beta}
\right) dx^\mu dx_\mu + \theta_{\mu\alpha} {F^\alpha}_\nu dx^\mu
dx^\nu = 0.
\ee
If we consider the case where there is no noncommutativity between
space and time, that is $\theta^{0i}=0$, and
calling $\theta^{ij} = \epsilon^{ijk} \theta^k$, $F^{i0} = E^i$, and
$F^{ij} = \epsilon^{ijk} B^k$, we find to first order in $\theta$ that
\be
\label{photon_eq}
( 1 -  \vec{v}^2 )( 1 - 2 \vec{\theta} \cdot \vec{B}) - \vec{\theta}
\cdot (\vec{v} \times \vec{E} ) + \vec{v}^2 \vec{\theta} \cdot \vec{B}
- (\vec{B} \cdot \vec{v}) ( \vec{\theta} \cdot \vec{v}) = 0,
\ee
where $\vec{v}$ is the particle velocity. Then to zeroth order, the
velocity $\vec{v}_0$ satisfies
$\vec{v}_0^2 = 1$ as it should. We can now decompose all vectors into
their transversal and longitudinal components with respect to
$\vec{v}_0$, $\vec{E} = \vec{E}_T + \vec{v}_0 E_L$, $\vec{B} =
\vec{B}_T + \vec{v}_0 B_L$ and $\vec{\theta} = \vec{\theta}_T + \vec{v}_0
\theta_L$. We then find that the velocity is
\be
\label{velocity_complex_scalar}
\vec{v}^2 = 1 + \vec{\theta}_T \cdot ( \vec{B}_T - \vec{v}_0 \times
\vec{E}_T ).
\ee
Hence, a charged massless particle has its velocity changed with
respect to the velocity of light by an amount which depends on
$\theta$. For an uncharged massless particle
\be
\label{velocity_real_scalar}
\vec{v}^2 = 1 + 2 \vec{\theta}_T \cdot ( \vec{B}_T - \vec{v}_0 \times
\vec{E}_T ),
\ee
and the correction due to the noncommutativity is twice that of a charged
particle.

We can now check the consistency of these results by going back to the
original action (\ref{action_scalar_field}) and
computing the group velocity for planes waves. Upon quantization
they give the velocity of the particle associated to the
respective field. For the uncharged scalar field we get the equation
of motion
\be
\label{eq_motion_real_scalar}
\left( 1 - \frac{1}{2} \theta^{\mu\nu} F_{\mu\nu} \right) \Box
\varphi - 2 \theta^{\mu\alpha} {F_\alpha}^\nu \partial_\mu\partial_\nu
\varphi = 0.
\ee
If the field strength is constant we can find a plane wave solution
with the following dispersion relation
\be
\label{dispersion_relation_real_scalar}
\left(  1 - \frac{1}{2} \theta^{\mu\nu} F_{\mu\nu} \right) k^2 - 2
\theta^{\mu\alpha} {F_\alpha}^\nu k_\mu k_\nu = 0,
\ee
and using the same conventions for vectors as before, it results in
\be
\label{dispersion_relation_vector_form}
\frac{\vec{k}^2}{\omega^2} = 1 - 2 \vec{\theta}_T \cdot ( \vec{B}_T -
\frac{\vec{k}}{\omega} \times \vec{E}_T ),
\ee
where $k^\mu = (\omega, \vec{k})$.
We then find that the phase and group velocities coincide and are given
by (\ref{velocity_real_scalar}) as expected. For the charged scalar
field we have to turn off the gauge coupling in order to get a plane
wave solution. In this case the equation of motion is
\be
\label{eq_motion_complex}
\left( 1 - \frac{1}{4} \theta^{\mu\nu} F_{\mu\nu} \right) \Box
\phi - \theta^{\mu\alpha} {F_\alpha}^\nu \partial_\mu\partial_\nu
\phi = 0.
\ee
In a constant field strength background the dispersion relation for a
plane wave reads as in (\ref{dispersion_relation_real_scalar}) with
$\theta$ replaced by $\theta/2$. Then we must perform the same
replacement in the phase and group velocities and we get
(\ref{velocity_complex_scalar}). Therefore, in both pictures,
noncommutative and gravitational, we get the same results.

For the gauge field the situation is more subtle because of its double
role. There is no clear way to
split the action (\ref{SW_gauge_action}). What can be done is to
break up the gauge field into a background plus a plane wave as 
in \cite{Guralnik:2001ax}. We then get the following dispersion
relation
\be
\label{dispersion_relation_gauge}
k^2 - 2 \theta^{\mu\alpha} {F_\alpha}^\nu k_\mu k_\nu = 0,
\ee
where ${F_\alpha}^\nu$ is now the constant background. This
leads to (\ref{dispersion_relation_vector_form}), that is, the
dispersion relation for the uncharged scalar field. It also reproduces
the result in \cite{Guralnik:2001ax,Cai:2001az} when the background is purely
magnetic. This shows the dual role of the gauge field, since it
couples to gravitation as a charged field but its dispersion relation
is that of an uncharged field.

We have seen that it is possible to regard
noncommutative theories as conventional theories embedded in a
gravitational background produced by the gauge field. This brings a
new connection between noncommutativity and gravitation. We could
imagine that this is a peculiarity of the first order term in the
$\theta$ expansion of the SW map but an analysis to all orders in
$\theta$ was performed in \cite{Banerjee:2004rs}. 

\vspace{1cm}

I would like to thank the organizers for the kind invitation to
deliver this talk. 
This work was partially supported by FAPESP, CNPq and PRONEX
under contract CNPq 66.2002/1998-99. 

\samepage

\end{document}